\documentclass[sn-mathphys,Numbered]{sn-jnl}
\usepackage{graphicx}%
\usepackage{multirow}%
\usepackage{amsmath,amssymb,amsfonts}%
\usepackage{amsthm}%
\usepackage{mathrsfs}%
\usepackage[title]{appendix}%
\usepackage{xcolor}%
\usepackage{textcomp}%
\usepackage{manyfoot}%
\usepackage{booktabs}%
\usepackage{algorithm}%
\usepackage{algorithmicx}%
\usepackage{algpseudocode}%
\usepackage{listings}%
\usepackage{lineno}

\begin{document}

\title[Article Title]{Multiplexing approaches to thermoradiative signatureless communications}

\author[1,2]{\fnm{Valerii} \sur{Radchenkov}}

\author[2]{\fnm{Peter J.} \sur{Reece}}

\author[1]{\fnm{Stephen P.} \sur{Bremner}}

\author[1]{\fnm{Nicholas J.} \sur{Ekins-Daukes}}

\author*[1]{\fnm{Michael P.} \sur{Nielsen}}\email{michael.nielsen@unsw.edu.au}

\affil[1]{\orgdiv{School of Photovoltaic and Renewable Engineering}, \orgname{UNSW Sydney}, \orgaddress{\city{Kensington}, \postcode{2052}, \state{NSW}, \country{Australia}}}

\affil[2]{\orgdiv{School of Physics}, \orgname{UNSW Sydney}, \orgaddress{\city{Kensington}, \postcode{2052}, \state{NSW}, \country{Australia}}}

\abstract{Using mid-infrared emission from  semiconductor devices for covert communications remains a relatively unexplored and yet promising opportunity. The phenomenon of negative luminescence allows for a method of signatureless covert communications where the net infrared emission of an emitting optoelectronic device is balanced to be identical to the ambient thermal background. In this work we provide a practical demonstration of covert data transfer over a thermoradiative channel with data rates up to 100 kbps. In addition, we demonstrate several additional multiplexing techniques that make the proposed thermoradiative communications method more secure against interception by achieving zero instantaneous optical emission, while remaining detectable if a sufficiently spatially or spectrally discerning observation is utilised. Finally, we discuss various application scenarios in which the proposed methods can be used to achieve secure signatureless communications.}

\keywords{mid-infrared communication, thermoradiative diodes, negative luminescence}

\maketitle

\section{Introduction}\label{sec1}

In applications where the security of communications is important, it is common practice to encrypt the communication channel or otherwise protect the message from being read if it is intercepted by a third party observer. However, a more secure way of preventing interception is to hide the very act of communication and the fact that there is any message being transmitted at all, such as by hiding transmission in noise for instance \cite{bash2015hiding}. Optical communication in the mid-infrared part of the spectrum presents a unique method of covert communications where the transmitter can be made optically indistinguishable from thermal background \cite{Nielsen2026}, hiding the presence of any communication. This is accomplished by rapidly emitting and absorbing infrared radiation, making the emitter look effectively "warmer" or "colder", and has previously been accomplished through efforts such as the electrocaloric effect \cite{bo2024flexible} or the electro-thermo-optic effect \cite{weinstein2022} or physical temperature changes \cite{guerra2020}. Using semiconductor devices as mid-IR emitters makes this possible due to the phenomenon of negative luminescence (NL), where a semiconductor photovoltaic device emits less light than in thermal equilibrium if a reverse bias is applied to it \cite{Berdahl1989, ashley:95, elliott2001, smith2007}. Unlike typical optical communications where only forward biased electroluminescence (EL) or other ``bright" emission is used for data transfer, using mid-infrared thermoradiative devices makes it possible to make the emitter look alternately ``hotter" or ``colder" relative to its environment by changing the sign of its bias voltage. As this effective carrier temperature can be modulated rapidly without changing the lattice temperature, it creates the possibility of using both positive and negative luminescence to carry information. This allows hiding the emission signature of the transmission by making the emitter look the same temperature as its surroundings. Such a method of covert communication (further referred to as thermoradiative signatureless communication \cite{Nielsen2026}) can be implemented by sending data through alternating EL and NL pulses such that the time-average emission is identical to that of the device's surroundings for a sufficiently slow detector or observer. 

Building upon the previous demonstration of thermoradiave signatureless communication \cite{Nielsen2026}, this work demonstrates a practical application of such a communication channel to transmit digital data at rates up to $\approx$ 100 kbps while maintaining an average infrared signature similar to thermal background, with transfer rates limited only by the available detector setup. In addition, we assess the feasibility of methods of further improving the security of a thermoradiative communications system against broadband fast infrared detection. This includes multiplexing the signal, either spatially or spectrally, to maintain thermal sampling neutrality. Here we explore the use of several closely-spaced emitters for scrambling or canceling each other's emissions as well as instantaneous masking of NL/EL emission using multiple emitters at different wavelengths.
\section{Demonstrating Thermoradiative Signatureless Data Transfer}\label{sec2}

\begin{figure}[H]
\centering
\includegraphics[width=1\textwidth]{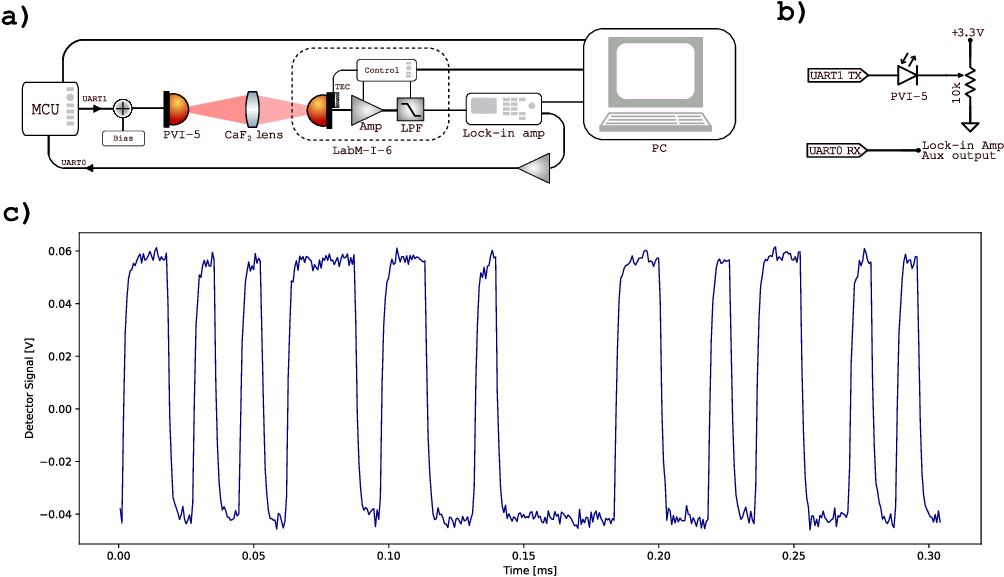}
\caption{(a) Schematic representation of the experimental setup. The RP2040 MCU transmits a UART signal through the PVI-5 photodiode, the signal is then received at the PVI-4TE-6 photodetector and passed through a lock-in amplifier. The lock-in amplifier filters and offsets the signal to the correct digital voltage levels, before sending it back to the MCU for detection. (b) Circuit schematic of the wiring of the PVI-5 emitter and the MCU. The device bias is adjusted to achieve correct EL and NL magnitudes to mask the signal. (c) The DC-coupled signal at the lock-in amplifier input, showing the UART waveform. The signal looks like it has a DC offset due to the MCU sending more 0s than 1s on average, resulting in a time-averaged null signal.}\label{fig1}
\end{figure}

Here we provide a practical demonstration of thermoradiative signatureless communication with a transfer of data over UART protocol, with measurements of attainable bandwidth and error rate. The experimental setup is depicted in Figure \ref{fig1}a. The demonstration setup uses a microcontroller (MCU) RP2040, in the form of a Raspberry Pi Pico development board to send a randomly-generated data packet over a UART channel. The UART output of the MCU is used to drive a MIR photodiode (Vigo Photonics PVI-5) that nominally emits at 5$\mu$m at the edge of the atmospheric transmission window. The photodiode is biased relative to the supply voltage of the MCU, such that the total emission of the device on average is close to thermal background (Figure \ref{fig1}b). The emitted infrared signal is then focused onto a Vigo Photonics PVI-4TE-6 photodetector with a nominal 6$\mu$m cut-off that is mounted inside a PIP-series electronics module (which includes a pre-amplifier, filters, device bias and a temperature controller). This electronics module ultimately controls the frequency cut-off of the detector system. The output of the detector is then connected to a lock-in amplifier (Zurich Instruments HF2LI), which in this demonstration is used to record the signal. The lock-in amplifier in this setup is used to amplify and offset the signal before it is sent back to the MCU, so that the signal is in the correct logical voltage range. The transmission and reception of data is done by two independent MCU cores as well as through different UART channels (UART1 for transmission, UART0 for reception). The received data frame is then compared to the original and the error rate is calculated as the proportion of incorrect bits.

The presented setup was used to transmit a frame of $10^{5}$ ASCII characters over the signatureless thermoradiative channel at a rate of up to 115200 baud (92.16 kbps) with error rate below $10^{-6}$. The exact measurement of error rate could not be performed due to memory limitations of the MCU, with no errors out of the $10^5$ characters that were sent. Higher baud rates could not be achieved due to the bandwidth limit of the lock-in auxiliary output (200 kHz), indicating that higher modulations rates and data transfer rates are possible with a different detector setup. The bias of the transmitter photodiode was adjusted such that the net emission as measured by the lock-in was identical to thermal background.

Figure \ref{fig1}c shows a fragment of the data frame as captured by the lock-in. On the graph the thermal background and detector bias are already subtracted, with a 0V detector signal corresponding to the thermal background. It can be observed that the positive half (EL mode) of the signal is approximately equal to the negative half (NL mode), but not precisely. The slight offset is due to the choice of transmitted data: printable ASCII character set does not cover the whole range of possible 8-bit sequences, resulting in a slight bias towards zeros in the transmitted frame. This means that the slight offset of EL over NL is used to counteract the higher number of ``0"s compared to ``1"s in the transmitted signal. The average detector voltage over the whole captured data frame was found to be around ~1.18 mV, or around 2\% of the peak detector voltage, and could be further compensated for with a different communications protocol designed for thermoradiative signatureless communications. This discrepancy is slightly higher than the thermal background, most likely due to incomplete detector bias cancellation over the short duration of the transmitted frame.

\section{Frequency Mutliplexing in a Signatureless Thermoradiative Channel}\label{sec3}

The signatureless communication method described above can still be intercepted if a sufficiently fast broadband infrared detector is used. To counteract this, additional techniques must be utilised to further disguise the act of communication, in this case through multiplexing arrangements. One method is to transmit several different signals through multiple spatially close emitters. By transmitting noise or random bits through some of the emitters and modulating the others at different frequencies, the transmitted signal would be further hidden from detection. In addition, several emitters can be configured to transmit the same signal out-of-phase, instantaneously canceling out each other's emission in the far field, which would require a detector with very high spatial resolution to intercept it.

\begin{figure}[H]
\centering
\includegraphics[width=1\textwidth]{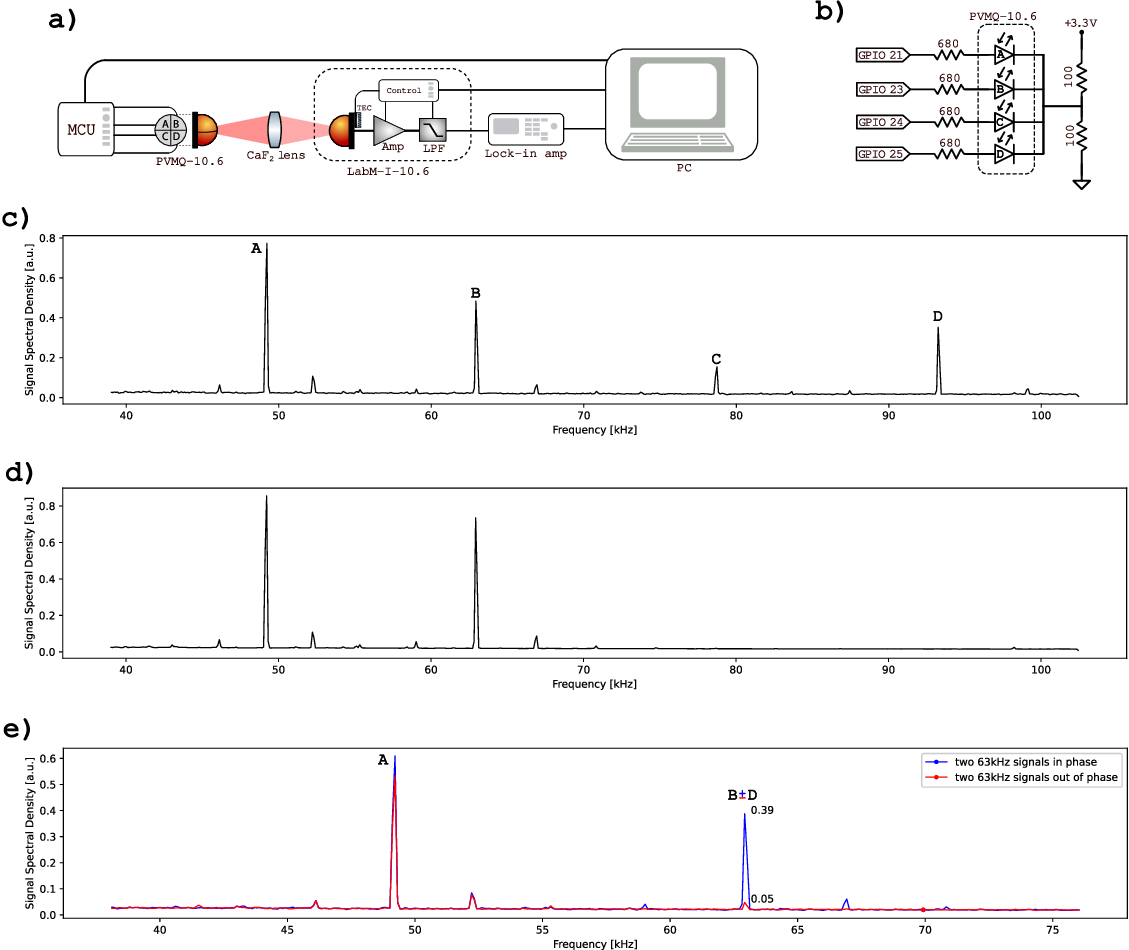}
\caption{(a) Schematic representation of the experimental setup. The four-quadrant device (PVMQ-10.6) is connected to the MCU which sends several frequency-separated signals. The photodetector passes the total signal to the lock-in amplifier, which can tune into the correct frequency of each of the four emitters. (b) Circuit schematic of the wiring of the PVMQ-10.6 device and the MCU, each device is biased to cancel its own emission in EL and NL modes.
(c) Spectrum of the signal in the input of the lock-in amplifier, with the frequency peaks corresponding to each of the device quadrant emitters (the left-most peak is environmental interference). (d) Same as (c), but a single device is emitting at two frequencies simultaneously. (e) Same as (c), but two of the quadrant emitters are fed at the same carrier frequency with a phase difference of 0 (blue) or 180 (red) degrees.}\label{fig2}
\end{figure}

Those two approaches are demonstrated using the setup in Figure \ref{fig2}a. An infrared device (Vigo Photonics PVMQ-10.6) with four quadrant detectors spaced closely together is used to transmit four different signals modulated at different frequencies. The MCU (RP2040) is configured to send several continuous square waves, with the signals generated simultaneously by frequency-dividing the MCU's system clock. The infrared emission of the four-quadrant device (Vigo Photonics PVMQ-10.6) is focused onto a Vigo Photonics PVI-4TE-10.6 photodetector incorporated into an electronics module (with temperature control and a pre-amplifier), which is connected to the lock-in amplifier/oscilloscope as before. The lock-in mixer can be tuned to each of the transmitting frequencies to detect the corresponding signals (in this case, unmodulated square wave carriers). The spectrum of the signal on the lock-in input is shown in Figure \ref{fig2}c. The peaks A-D correspond to the four device quadrants, with different detection sensitivities given the setup limitations. Additional peaks arise in this case from background noise. Each of the signals alone is on average indistinguishable from thermal background as emission in EL mode is canceled out by ``dark beam'' emission in NL mode. The biasing circuit used to accomplish this is shown in Figure \ref{fig2}b. The emission intensity in both NL and EL modes for the given driving voltage is very similar for this device due to the narrow bandgap \cite{Nielsen2026}.

Figure \ref{fig2}d shows the same communication method implemented using a single emitter, in this case 'A'. Here, two signals are fed into the opposite terminals of the photodiode at different frequencies, such that the total infrared emission is the difference of the two signals. This scrambles the signals in time domain (if one of the signals is random bits), but allows the message to be detected if the correct carrier frequency is used.

It is also possible to configure two of the emitters in the four-quadrant device to emit the same signal but 180 degrees out of phase, which would lead to them canceling each other out in the far field if there is insufficient imaging resolution from an observer. Figure \ref{fig2}e shows the signal spectrum on the lock-in input for such a configuration: the emitter A is continuously transmitting at one frequency, while B and D transmit on another frequency with the two emitters having varying phases. On the blue trace, the two emitters are in phase, reinforcing each other, while on the red trace the signals are 180 degrees out of phase resulting in them masking each other (the masking is imperfect due to setup limitations). Detecting such a signal would require a photodetector with very high spatial resolution to resolve the individual device quadrants. If the signal is transmitted via an MIR optical fiber, the different device quadrants may be used to excite different optical modes of the fiber such that two modes cancel each other when the intensity is averaged along the fiber's cross-section, but the signal can still be detected if the receiver can distinguish the two modes.

\section{Hiding Instantaneous Emission Using Wavelength Multiplexing}\label{sec4}

Another method of further concealing the described signatureless communication channel is to require the receiver to be wavelength-selective. If two spatially and spectrally close, but not wavelength overlapping, emitters were to provide the same data signal but at reversed polarity this could be tuned to look exactly like the thermal background even to an observer with a fast broadband detector unless they have the correct narrowband filter to distinguish the signal. In this case both the spectrally-averaged time-resolved signature and the time-averaged spectrally resolved signature would be zero, with only the time and spectrally resolved signal showing as non-zero to an observer. 

To demonstrate this, we used two devices emitting at different wavelengths (PVI-4 at 4$\mu$m and PVI-5 at 5$\mu$m) with phases and emission intensities matched, so that the two devices cancel out each other's emission instantaneously at the detector. This way the combined infrared emission is ideally equal to thermal background at all times. And yet, when observing the emitter through a wavelength filter that passes the emission of only one device but not the other would then reveal the transmitted signal.  

\begin{figure}[H]
\centering
\includegraphics[width=1\textwidth]{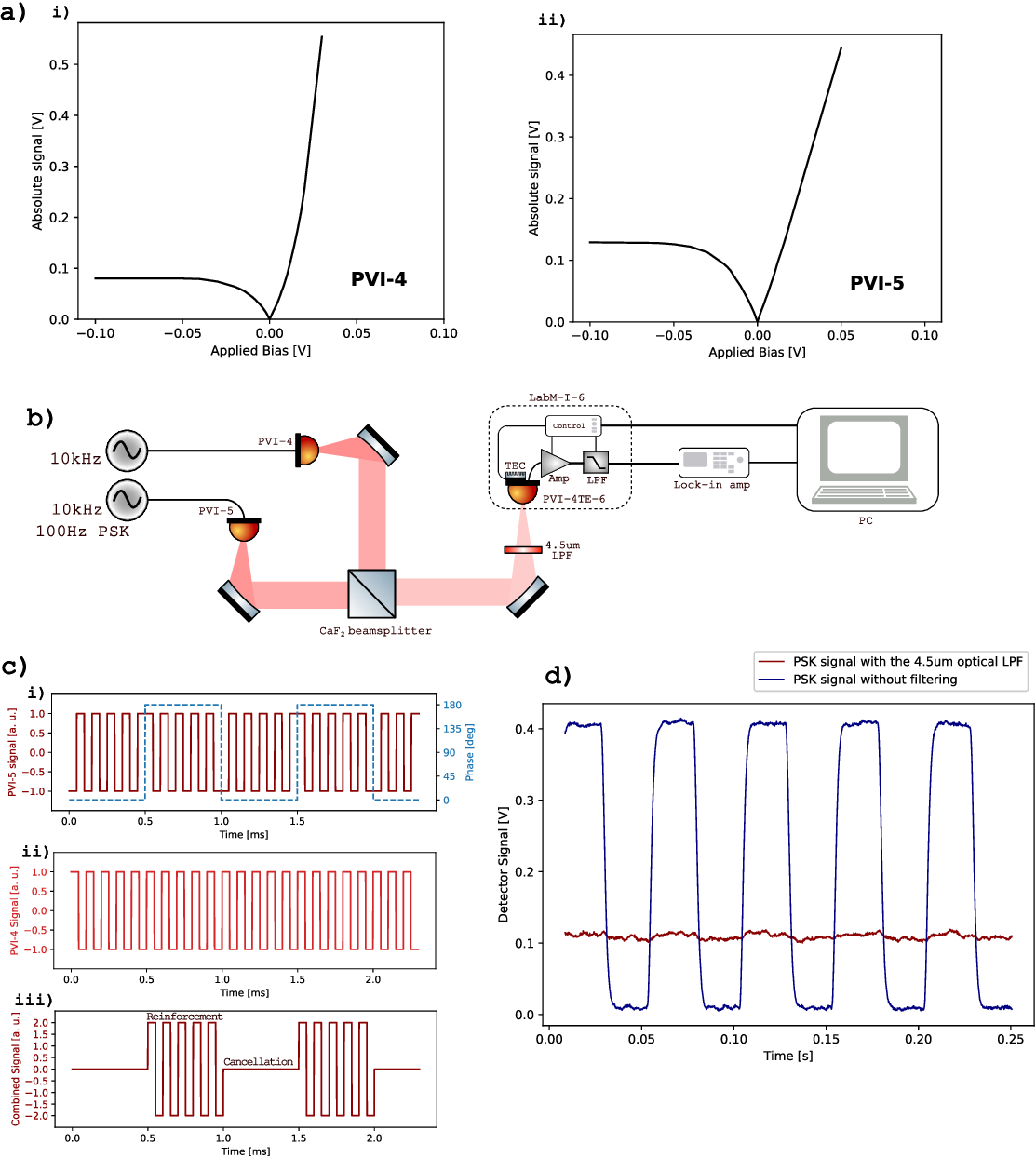}
\caption{(a) Detected emission from the PVI-4 (i) and PVI-5 (ii) photodiodes as a function of bias voltage at 10kHz (50\% duty cycle) in both NL and PL modes. (b) Schematic representation of the
experimental setup. The emission of the two devices is combined at the photodetector (PVI-4TE-6) using a beamsplitter, and the resulting signal is recorded at the lock-in amplifier.
A removable 4.5um optical low-pass filter is added in front of the photodetector to allow discrimination between devices.
(c) Conceptual explanation of the device phasing: (i) a phase-shift-keyed (PSK) 10kHz square wave is passed to the PVI-5 emitter; (ii) an unmodulated 10kHz square wave is passed to the PVI-4 emitter;
(iii) the combined signal as recorded by the photodetector. (d) The recorded signal on the output of the lock-in amplifier, after mixing from 10kHz. Without a 4.5um optical LPF (blue curve), the received 
signal is cancelled and set to zero when the devices have opposite phase but are reinforced when in phase. Adding the 4.5um LPF (red curve) allows the signal from the PVI-5 device to be detected as a constant signal over this time period.}\label{fig3}
\end{figure}

In order for this method of covert communication to work well, the magnitude of EL and NL emissions of the two devices must be closely matched, with the EL emission of each device adjusted to match the saturated NL of the other device. This is because the peak emission of each device needs to be as high as possible for optimal noise performance, but the magnitude of negative luminescence scales with wavelength and exhibits a saturation behaviour. Figure \ref{fig3}a shows the recorded photodetector signal for the two devices as function of applied bias, which was used to set the appropriate DC offsets for the two devices. In the setup, both devices were driven until saturation in the NL mode with a 100mV negative half-wave, and the positive half-wave was adjusted to cancel the NL emission of the other device.

In the experimental setup shown in Figure \ref{fig3}b, the infrared emission from the two devices is combined using a system of parabolic mirrors and a beamsplitter, and then imaged onto the photodetector module with an integral PVI-4TE-6 device. Then, by adding a 4.5um low-pass optical filter, the individual emission from the 5um device alone is revealed. Both devices were configured to transmit a 10kHz square wave at 50\% duty cycle, with DC offsets adjusted so that the emission of one device in EL mode completely cancels the emission of the other in NL mode, and vice versa. To demonstrate the instantaneous signal cancellation for the two devices at different wavelengths, one transmitted a continuous square wave (Figure \ref{fig3}c(ii)), while the other had its phase modulated between 0 and 180 degrees (phase-shift keying, or PSK) at 100Hz, as in Figure \ref{fig3}c(i). The combined signal at the photodetector in this configuration would show periodic cancellation and reinforcement of emission between the two devices (Figure \ref{fig3}c(iii)). In a real-world implementation the two devices would be configured to emit out-of-phase all of the time, and using PSK in this setup is done purely for demonstration purposes to show that the amplitude variation disappears when an optical filter is inserted before the detector. After the lock-in mixes the signal with the 10kHz carrier, only the modulating PSK waveform would be visible, showing the relative magnitudes of the signal when the two devices cancel or reinforce each other. When the 4.5$\mu$m optical LPF is added, the PSK waveform disappears and only the continuous signal (over this time frame) from one device (PVI-5) remains (Figure \ref{fig3}d. It is worth noting that in this demonstration the optical filter is not perfectly selective, as evident by the slight 100Hz oscillation in the red trace on Figure \ref{fig3}d, but could be engineered for tighter tolerances via narrower emission spectra or more discerning wavelength filtering.

A practical application to this method of covert communication would require the intended receiver of the signal to know the exact wavelength at which the message is transmitted and to be able to discriminate between them. A more advanced version of this method could entail using several pairs of emitters taking up a band somewhere in the atmospheric transmission window. Different signals could be encoded in different parts of the band, requiring a corresponding optical band-pass filter on the receiver side, with a further algorithm to move between wavelength channels in a pre-determined manner. In this way one transmitter could be used to emit more than one message where each is targeted at one receiving party neither of which has access to the other transmitted messages. To a broadband receiver trying to spy on the communication channel the whole transmission would look identical to background thermal radiation.

\section{Future Directions}\label{sec5}

While promising, the proposed thermoradiative covert communication technology requires several important improvements to be practical. One of the current limitations is the range at which the data transfer is effective, being restricted by saturation in negative luminescence as well as limited directivity of the devices used in the demonstration. While negative luminescence has an ultimate lower bound of approaching effectively 0K optical temperature, to date experiental demonstrations have not been able to approach this level of temperature suppression \cite{lindle2006hgcdte}. Several quantitative advances in transmitted power, directivity and bitrate are possible through improving the parameters of the emitter and photodetector.

The directivity of the devices used in the experimental setup is primarily limited by the use of an immersion lens, which also imposes a limit on the emitter area. Using metasurfaces \cite{cortes2022optical, chu2024controlling, cortes2022optical} or metalenses \cite{Bogh2020, Zhang2018, Wang2019b} to focus the beam allows for improved directivity. In addition, resonant metasurfaces provide much more narrowband emission than the original device, which could offer a significant improvement for the wavelength multiplexing technique described above. Making use of certain 2D materials (black phosphorus \cite{youngblood2015waveguide}, graphene \cite{shautsova2018plasmon, gabor2011hot}) would allow for significantly improved bandwidth and data rates, with bandwidths up to 500 GHz demonstarted in graphene \cite{koepfli2023metamaterial} This could potentially allow for data bitrates comparable to current optical communication technologies.

The signal intensity in the proposed communication system is primarily limited by saturation in NL mode, which can be improved by using a lower-bandgap emitter. However, at low bandgaps non-radiative recombination mechanisms become more prevalent \cite{nielsen2024semiconductor,radchenkov2025temperature} which significantly limits efficiency and signal intensity. Addressing this would require developing materials with lower non-radiative recombination rates, or developing more efficient structures with existing materials, such as through the use of superlattice structures or multi-stage devices \cite{nielsen2024semiconductor}. Multi-stage devices would also allow for spatial stacking of devices with different emission wavelengths. The device structures that allow for efficient thermoradiative power generation \cite{Nielsen2022} would be also useful as an NL emitter with appropriate index matching as both the thermoradiative effect and negative luminescence require low non-radiative recombination rates in order to be efficient.

Additionally, the issue of power saturation can be partially mitigated by using MIR optical fibers for signal transmission. This would also solve the problem of limited emitter directivity (the emitter is practically a non-coherent thermal source) as well as concerns about atmospheric absorption at MIR wavelengths \cite{harrison2024evaluating}. Using a multi-mode optical fiber would allow one to use spatial (mode) multiplexing, with several devices masking each other in far-field, i.e. through two or more closely-spaced emitters transmitting the same message out-of-phase relative to each other. Each device would excite a different mode in the fiber, with the superposition of several out-of-phase device emissions instantaneously averaging to zero across the fiber cross-section. This would make it nearly impossible to detect the message from outside the fiber. However, a detector with sufficient spatial resolution would be able to discriminate between the optical modes and recover the signal.

Another possible application would make use of wavelength multiplexing approach. Emitters with metasurface focusing could be used to emit directional beams at a specific wavelength, with pairs of emitters canceling each other's emission instantaneously so that the whole transmission looks indistinguishable from thermal background if the intercepting detector cannot differentiate between the different emitter wavelengths. Each emitter pair could target a specific receiving device that is exclusively aware of its particular pair of wavelengths and can differentiate between them, allowing it to recover the message. Such a communication method would be secure against not only external spying but also accidental information leaks between different parties on the same network. As before, this kind of communication technology would be more practical to implement in an optical fiber, and it could even be combined with the mode-selective approach discussed above.
\section{Conclusions}\label{sec6}
In this work we have presented a practical demonstration of data transfer using thermoradiative signatureless communication channel, achieving rates up to 100kbps with no data loss. In addition, we have discussed and demonstrated several additional approaches to covert communications that build on our previous demonstration of thermoradiative signatureless communications, namely using different transmission frequencies, masking emitters via spatial resolution and wavelength multiplexing the emitters to mask each other instantaneously. Several limitations and possible directions for improvement were identified, such as modifying device materials for improved bit rate and NL mode emission, using metasurfaces to improve directivity and frequency selectivity. Lastly, we have presented several potential application scenarios of how the proposed thermoradiative data transfer can be used in a practical communication network.

\backmatter


\subsection*{Declarations}
\bmhead{Data Availability}
Data is available from the authors upon reasonable request.

\bmhead{Competing interests}
M.P.N. and N.J.E-D have filed a patent application (Australian Provisional Patent Application No. 2025904829) related to the methods reported in this paper.

\bmhead{Contributions}
M.P.N. proposed the project. All authours discussed the implementation of the concept. V.R. performed the experiments and data analysis. V.R. and M.P.N. wrote the manuscript with input from all authours.

\bmhead{Acknowledgments}
M.P.N. recognises the support of the UNSW Scientia Program and a Spitfire Memorial Defence Fellowship. 

\bibliography{references}

\end{document}